\documentclass[english,prl,preprintnumbers]{revtex4}
\usepackage[T1]{fontenc}
\usepackage[latin9]{inputenc}
\usepackage{amstext}

\makeatletter
\@ifundefined{textcolor}{}
{%
 \definecolor{BLACK}{gray}{0}
 \definecolor{WHITE}{gray}{1}
 \definecolor{RED}{rgb}{1,0,0}
 \definecolor{GREEN}{rgb}{0,1,0}
 \definecolor{BLUE}{rgb}{0,0,1}
 \definecolor{CYAN}{cmyk}{1,0,0,0}
 \definecolor{MAGENTA}{cmyk}{0,1,0,0}
 \definecolor{YELLOW}{cmyk}{0,0,1,0}
 }

\makeatother

\usepackage{babel}

\begin{document}

\preprint{MIT-CTP-4182}

\title{An online attack against Wiesner's quantum money}

\author{Andrew Lutomirski}

\affiliation{Center for Theoretical Physics, Massachusetts Institute of Technology,
Cambridge, MA 02139}

\email{luto@mit.edu}

\date{September 31, 2010}
\begin{abstract}
Wiesner's quantum money \cite{Wie83} is a simple, information-theoretically
secure quantum cryptographic protocol. In his protocol, a mint issues
quantum bills and anyone can query the mint to authenticate a bill.
If the mint returns bogus bills when it is asked to authenticate them,
then the protocol can be broken in linear time.
\end{abstract}
\maketitle

\section{Introduction}

In \cite{Wie83}, Wiesner proposed a protocol for information-theoretically
secure private-key quantum money. A mint can choose a security parameter
$n$ and generate a random $n$-qubit state. (Each qubit is independently
and uniformly drawn from $\left\{ |0\rangle,|1\rangle,|+\rangle,|-\rangle\right\} $.)
The mint assigns that state a unique serial number and declares it
to be a \$20 quantum bill. (The \$20 is arbitrary.) To verify a quantum
bill, a merchant sends the bill to back to the mint. The mint looks
up the classical description of the state matching the serial number
of the bill and projects the quantum state being tested onto that
state. A {}``\textbf{VALID}'' result (the state being tested matched
the description) means that the bill is valid and an {}``\textbf{INVALID}''
result means that the bill was counterfeit or damaged. (The mint needs
to maintain a secret database of the description of the random state
corresponding to each serial number.)

This protocol is information-theoretically secure. The no-cloning
theorem implies that an attacker cannot perfectly copy a quantum bill,
and the bounds in \cite{buzek1996quantum} mean that the probability
that an approximate copy appears valid drop exponentially as a function
of $n$.

There are many recent papers based on the idea of attacking otherwise
secure \emph{classical} cryptographic protocols by various side channels
or online attacks. For example, in 2002, Vaudenay showed that a commonly
used form of symmetric cipher (CBC mode encryption) can be attacked
with a small number of queries to an oracle that distinguishes valid
encrypted messages from invalid messages with a certain type of error
\cite{vaudenay2002security}. Rizzo and Duong dramatically showed
that these attacks worked against carelessly designed websites and
that many current websites are vulnerable \cite{rizzo2010padding}.

Inspired by Rizzo and Duong's result, I show that, even if the mint
has a perfect quantum computer, Wiesner's quantum money is vulnerable
to an online attack. If, when asked to verify any bill, the mint returns
the bill \emph{even if that bill was invalid}, then a small number
of queries to the mint can be used to copy a bill.

\section{The attack}

For Wiesner's quantum money to be useful, the mint must offer a service
that anyone can use to verify quantum bills. Morris (presumably a
merchant) sends the mint a quantum bill. The mint either answers \textbf{VALID}
and returns the bill to Morris or answers \textbf{INVALID}. In the
\textbf{INVALID} case, if the mint destroys the counterfeit bill,
then all is well. If, on the other hand, the mint returns the counterfeit
bill to Morris, then the entire protocol can be broken in linear time.

We can formalize the quantum bill as a classical-quantum state $\left(s,|\$_{s}\rangle\right)$
where $s$ is some unique classical serial number and $|\$_{s}\rangle$
is the random product state chosen by the mint that corresponds to
the serial number $s$. We can write \[
|\$_{s}\rangle=|\psi_{1}\rangle|\psi_{2}\rangle\ldots|\psi_{n}\rangle,\]
 where each $|\psi_{i}\rangle$ depends on $s$ and is drawn from
$\left\{ |0\rangle,|1\rangle,|+\rangle,|-\rangle\right\} $. This
means that each $|\psi_{i}\rangle$ is an eigenstate of either $X$
or $Z$. (Morris does not know which operator each $|\psi_{i}\rangle$
is an eigenstate of.) We assume that Morris can send the mint any
c-q state $\left(s,|\phi\rangle\right)$ and the mint will measure
the projector $P_{s}=|\$_{s}\rangle\langle\$_{s}|$. If the outcome
is 1, the mint returns $\left(\textbf{VALID},s,P_{s}|\phi\rangle\right)$
and if the outcome is 0, the mint returns $\left(\textbf{INVALID},s,\left(1-P_{s}\right)|\phi\rangle\right)$
(up to normalization).

If Carla the counterfeiter has a single quantum bill $\left(s,|\$_{s}\rangle\right)$
and can query the mint, then she can break the protocol by learning
the state $|\$_{s}\rangle$ one qubit at a time. To learn the $i$th
qubit, she sends the mint the state $\left(s,X_{i}|\$_{s}\rangle\right)$.
If the mint answers \textbf{INVALID}, then the state $|\psi_{i}\rangle$
was either $|0\rangle$ or $|1\rangle$ (as the other possibilities
$|+\rangle$ and $|-\rangle$ are eigenstates of $X_{i}$). In this
case, the returned state is \[
|\psi_{1}\rangle\cdots|\psi_{i-1}\rangle|\psi_{i}^{\perp}\rangle|\psi_{i+1}\rangle\cdots|\psi_{n}\rangle.\]
 But now Carla knows that $|\psi_{i}^{\perp}\rangle$ is an eigenstate
of $Z$. She applies $X_{i}$ to recover $|\$_{s}\rangle$ and measures
$|\psi_{i}\rangle$ in the $Z$ basis to learn whether it is $|0\rangle$
or $|1\rangle$.

If, on the other hand, the mint answers \textbf{VALID}, then the state
$|\psi_{i}\rangle$ was either $|-\rangle$ or $|+\rangle$. In this
case the mint returns the (undamaged) state $|\$_{s}\rangle$ to Carla.
But now Carla knows that $|\psi_{i}\rangle$ is an eigenstate of $X$
and she can measure it to learn whether it is $|+\rangle$ or $|-\rangle$.

If Carla repeats this process for $i=1,\ldots,n$, she will learn
the secret description of $|\$_{s}\rangle$ in exactly $n$ queries
to the mint. Once she has done this, she can make as many counterfeit
copies of $|\$_{s}\rangle$ as she wants.

Carla could also use a more generic algorithm such as quantum state
restoration to copy the state directly in $2n$ (expected) queries
to the mint or single-copy tomography to learn the state in $O\left(n\right)$
queries \cite{state-restoration}.

\section{Conclusion}

Anyone who implements classical cryptographic protocols needs to be
very careful to avoid introducing flaws that bypass the security that
the protocol would have offered if correctly implemented. Quantum
cryptography is not magically safer.

\section{Acknowledgments}

I was supported by the Department of Defense (DoD) through the National
Defense Science \& Engineering Graduate Fellowship (NDSEG) Program.

\bibliographystyle{hplain}
\bibliography{money-bib}

\end{document}